\documentstyle[prl,aps]{revtex}
\begin{document}
\include{epsf}

\title{Kondo Insulator: p-wave Bose Condensate of Excitons}

\draft

\twocolumn[\hsize\textwidth\columnwidth\hsize\csname @twocolumnfalse\endcsname

\author{Ji-Min Duan, Daniel P. Arovas, and L. J. Sham}
\address{Department of Physics, University of California
San Diego, La Jolla, California 92093-0319}

\date{\today}

\maketitle

\begin{abstract}
In the Anderson lattice model for a mixed-valent system, the
$d-f$ hybridization can possess a $p$-wave symmetry.  The
strongly-correlated insulating phase in the mean-field approximation is shown
to be a $p$-wave Bose condensate of excitons with a spontaneous lattice
deformation.   We study the equilibrium and linear response properties
across the insulator-metal transition.   Our theory supports the empirical
correlation between the lattice
deformation and the magnetic susceptibility and predicts measurable
ultrasonic and
high-frequency phonon behavior in mixed-valent semiconductors.
\end{abstract}

\vskip2pc]

\narrowtext

Mixed valent compounds containing $f$-electrons in their insulating phase
possess very distinctive electronic properties showing strong electron
correlation \cite{AeppliFisk,GangOf5}, and are commonly called Kondo
insulators, even though they are often not in the Kondo regime
\cite{Varma}.  The
insulating state may be regarded as a condensate of the slave bosons which are
used to simulate the high energy cost of too many
$f$-electrons on the same ion site
\cite{Millis}. Through the $f$-level and conduction band hybridization, this
Bose condensate drives the condensation of $f$-hole-conduction-electron
excitons, creating a band gap.  This theory has been used to explain a number of
thermal, transport, and optical properties \cite{Riseborough}.

In this paper, we adopt the same slave-boson treatment of the Anderson Lattice
Model to explore the consequences of the $d$-character of the conduction
electrons so that the $d$-$\!f$ hybridization at wave-vector ${\bf k}$
exhibits inversion antisymmetry, {\it i.e.}
\begin{eqnarray}
V(-{\bf k}) = - V({\bf k}) = V^*({\bf k}).
\label{hybrid}
\end{eqnarray}
This leads to odd-symmetry pairing of $d$-electrons and $f$-holes.
(See below on how a $p$-wave symmetry can arise.)
Coupling of the $d$-$\!f$ excitons to the lattice deformation
leads to a structural phase transition known as the ferroelastic transition
\cite{fet}. We deduce the temperature dependence of the equilibrium properties
(lattice distortion and phase diagrams) and of the linear responses (including
the magnetic susceptibility, elastic constants, phonon frequency and damping).
These properties show the characteristic dependence on the $p$-wave condensate
and their observation can be used to distinguish the $s$- or $p$-wave nature of
the Kondo insulator state.  We shall make the case below that some of the
$p$-wave characteristics have already been found in certain Kondo insulators.
The establishment of the $p$-wave characteristics in a class of Kondo
insulators could be of importance in providing a central framework for
understanding their unusual properties in terms of the symmetry of the
condensate.

The contrast between the $p$-wave condensate in the Anderson lattice and the
$s$-wave condensate in the Falicov-Kimball model \cite{Portengen} should also
be noted. In the latter, the $d$-$\!f$ hybridization is neglected in comparison
with the strong $d$-$\!f$ Coulomb attraction which is responsible for the
exciton
condensation.  Since each $d$-$\!f$ exciton carries a dipole moment, the
condensate behaves like an ``electronic'' ferroelectric.  On the other hand,
from symmetry considerations, the $p$-wave condensate does not produce a
macroscopic electric polarization but can yield a finite lattice distortion.
Responses to lattice vibrations induce coupling to the collective modes of the
$p$-condensate, which are fundamentally different from the electronic
ferroelectrics.  When both the hybridization
and $d$-$\!f$ interaction are present, the properties of the $s$ and 
$p$ wave coexist. Details will be presented in a long paper
while here we concentrate on the limiting case when hybridization
dominates, in contrast to the other limit \cite{Portengen}.

We consider a model with one $f$-electron and one conduction
electron per unit cell \cite{Millis}.  For simplicity, we include only the
one component each of the $d$ and $f$-bands which yield the hybridization
obeying
Eq.~(\ref{hybrid}) and neglect the $f$ orbital degeneracy.  To consider the
formation of the condensate and the related dynamic response, we replace the
onsite interaction in the Anderson model by the slave-boson contraints and
further retain only zero-momentum slave bosons which are present in macroscopic
numbers \cite{Millis}.  The Anderson Hamiltonian is thus reduced to:
\begin{eqnarray}
H
= \sum_{{\bf k}\sigma} \left[ \epsilon ({\bf k}) -\mu \right]
d^\dagger_{{\bf k}\sigma} d^{\vphantom\dagger}_{{\bf k}\sigma}
+ \left(\epsilon_f -\lambda -\mu \right) \sum_{{\bf k}\sigma}
 f^\dagger_{{\bf k}\sigma} f^{\vphantom\dagger}_{{\bf k}\sigma} \nonumber \\
+\frac{1}{\sqrt{N_{\rm s}}} \sum_{{\bf k}\sigma} \{  V({\bf k})\,
 d^\dagger_{{\bf k}\sigma} f^{\vphantom\dagger}_{{\bf k}\sigma}
b^\dagger + \text{h.c.}\}
+ \lambda \left( N_{\rm s} - b^\dagger b \right) ,
\label{hami}
\end{eqnarray}
where $d_{{\bf k}\sigma}, f_{{\bf k}\sigma}, b$ are annihilators for the
conduction and $f$ electrons of wave vector ${\bf k}$ and spin $\sigma=\pm 1$,
and slave-boson of zero momentum.  The conduction band energy is denoted by
 $\epsilon ({\bf k})$ and the chemical potential by $\mu$.  $\lambda$ is the
Lagrange multiplier which enforces the constraint of
at most one $f$-electron per cell for $N_s$ cells (on average).

The equations of motion for the $d$-$\!f$ exciton condensate are physically
transparent upon expressing the density matrix in terms of a pseudo-spin vector
${\bf S}_{\bf k}$:
\begin{eqnarray}
S^x_{\bf k} + i S^y_{\bf k} &=&
2\langle d^\dagger_{{\bf k}\sigma} f^{\vphantom\dagger}_{{\bf k}\sigma}\rangle
\nonumber \\
S^z_{\bf k} &=& \langle d^\dagger_{{\bf k}\sigma}
d^{\vphantom\dagger}_{{\bf k}\sigma}- f^\dagger_{{\bf k}\sigma}
f^{\vphantom\dagger}_{{\bf k}\sigma}\rangle.  \label{spin}
\end{eqnarray}
In the mean-field approximation, the spin vector satisfies a Bloch equation
\cite{Portengen}:
$\dot{\bf S}_{\bf k} = {\bf \Omega}_{\bf k} \times {\bf S}_{\bf k}$,
where the precession frequency is given by
\begin{eqnarray}
{\bf \Omega}_{\bf k} =  2\left( \Re\left[  V^*({\bf k})
\frac{\langle b \rangle}{\sqrt{N_{\rm s}}} \right],
\Im\left[  V^*({\bf k})
\frac{\langle b \rangle}{\sqrt{N_{\rm s}}} \right],
\xi({\bf k}) \right),
\label{angvel}
\end{eqnarray}
where $\Re$  and $\Im$ denote the real and imaginary parts and
$\xi({\bf k}) =\case{1}{2}
[\epsilon({\bf k}) - \epsilon_f +\lambda]$.
The expectation value at a finite temperature of the slave-boson condensate
$\langle b \rangle$ is driven by the exciton condensate\cite{foot}
$\langle d^\dagger_{{\bf k}\sigma} f^{\vphantom\dagger}_{{\bf k}\sigma}\rangle$:
\begin{eqnarray}
i \partial_t \langle b \rangle +
\lambda \langle b \rangle =  \frac{1}{\sqrt{N_{\rm s}}} \sum_{{\bf k}\sigma}
V({\bf k}) \langle d^\dagger_{{\bf k}\sigma}
f^{\vphantom\dagger}_{{\bf k}\sigma} \rangle.
\label{boso}
\end{eqnarray}

The zeroth-order (in the absence of external perturbation) solution of the above
equations yield the thermal equilibrium state.  The spin vector
${\bf S}^{(0)}_{\bf k}$ is a unit vector in the direction opposite to the zeroth
order precession frequency ${\bf \Omega}^{(0)}_{\bf k}$, yielding the exciton
order parameter:
\begin{eqnarray}
\langle d^\dagger_{{\bf k}\sigma} f^{\vphantom\dagger}_{{\bf k}\sigma}
\rangle^{(0)}= a V^*({\bf k}) \left[ f(E_{{\bf k}+})
- f(E_{{\bf k}-})\right]/2E_{\bf k},
\label{wvfn}
\end{eqnarray}
where $f(E)$ is the Fermi distribution, $a=\langle b\rangle^{(0)}/
\sqrt{N_{\rm s}}$ and the quasi-particle energies are
$E_{{\bf k}\pm}=  \case{1}{2} [\epsilon_f-\lambda^{(0)} +\epsilon ({\bf k})]
\pm E_{\bf k}$,
where $E_{\bf k} = \sqrt{[\xi^{(0)}({\bf k})]^2+
 a^2| V({\bf k})|^2 }$.
From Eq.~(\ref{wvfn}), the exciton order parameter being driven by the
hybridization $V({\bf k})$ must have the same symmetry.
By Eq.~(\ref{hybrid}), the pairing state is antisymmetric under
inversion.  More specifically, if $V({\bf k})$ is one of the $p$
components of the hybridization matrix elements between the d and f states,
the exciton order parameter would indeed have a $p$-wave symmetry.

To evaluate the mean field phase diagram from the zeroth order equations,
we further adopt a simple model with the conduction band having a constant
density of states with a bandwidth of $2W$.  The hybridization is approximated
by $V({\bf k})=i\,V_0\,\text{sgn}(k_z)$, where $V_0$ is a real constant.
The vanishing gap case is avoided by choosing the energy
surface $\xi^{(0)}({\bf k})=0$ not to intersect the plane
where $V({\bf k})$ vanishes. 
Some of the results are: (1) At $T=0$ for arbitrary $\epsilon_f$, the ground
state is always a $p$-wave condensate of excitons (correlated insulator).
(2) For $\epsilon_f \geq 0$, there is no finite-temperature phase transition.
(3) For $\epsilon_f < 0$, there is a second-order phase transition at finite
temperature between the correlated insulator and a simple metal. 

\begin{figure} [!t]
\centering
\leavevmode
\epsfxsize=7cm
\epsfysize=7cm
\epsfbox[18 144 592 718] {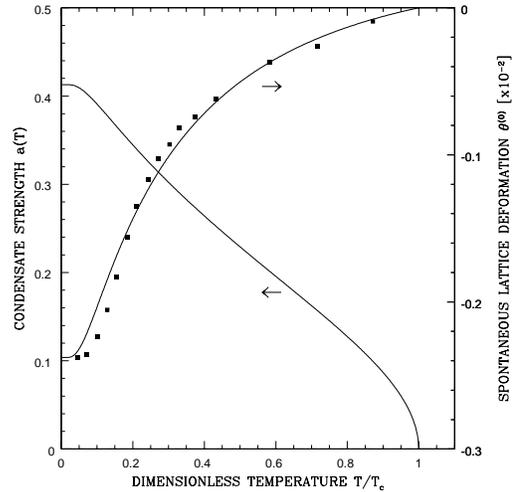}
\caption[]
{\label{fig1} Condensate strength and spontaneous lattice deformation versus
temperature for $\Xi_0 =-0.47 W$, $N_{\rm s}=3\times 10^{27}/m^3$, $C=
10^{10}J/m^3$ for the simple model whose other parameters are specified in the
text. Squares are experimental results from Ref.~10 for Ce$_3$Bi$_4$Pt$_3$.}
\end{figure}
Figure~1 shows the temperature dependence of the slave boson condensate $a(T)$
for $\epsilon_f=-0.2\, W$ and $V_0=0.3\, W$. The resultant $k_B T_c = 0.123\,W$
indicates this
case to be in the intermediate coupling region.
For the rest of this paper, the numerical results are reported for the same set
of parameters.

Two immediate consequences of the odd symmetry exciton condensate are the
absence of polarization and the possibility of lattice distortion.  For a more
quantitative study,  we include in the Anderson model a coupling with the strain
field $\theta$  through the deformation potential $\Xi({\bf k})$
\cite{ShamZiman} by replacing $V({\bf k})$ by $V({\bf k})+\Xi({\bf k})\,\theta$.
From the form of the electron-phonon interaction we have included, the
deformation potential $\Xi({\bf k})$ arises from the change of the
hybridization $V({\bf k})$ to first order in the lattice displacement and,
therefore, must have the same odd symmetry under inversion.
The lattice deformation creates a stress:
\begin{eqnarray}
s = C\theta + \frac{2}{\sqrt{N_{\rm s}}} \Re\sum_{{\bf k}\sigma} \Xi({\bf k})
\langle d^\dagger_{{\bf k}\sigma} f^{\vphantom\dagger}_{{\bf k}\sigma} \rangle
\langle b^\dagger \rangle ,
\label{deform}
\end{eqnarray}
where $C$ is the bare elastic constant.
The zeroth order terms show that the exciton condensate induces a {\it
spontaneous} lattice deformation $\theta^{(0)}$. The feedback of the lattice
distortion should be included in Eqs.~(\ref{angvel}) and (\ref{boso}).
A tight-binding derivation shows that $\Xi({\bf k})$
is the same order of magnitude as $V({\bf k})$ and the distortion
$\theta^{(0)}$ is always less than a percent (see below).
Therefore, we may neglect the self-consistent inclusion of the
lattice distortion in the determination of the exciton and slave-boson
condensates.

Figure~1 also shows the temperature dependence of the lattice
distortion $\theta^{(0)}$ in comparison with experiment on Ce$_3$Bi$_4$Pt$_3$,
chosen as an example of a class of small-gap correlated
semiconductors \cite{Fisk,AeppliFisk}. The deformation potential is approximated
in a similar way to the hybridization:
$\Xi({\bf k})=i\,\Xi_0\,\text{sgn}(k_z)$ where $\Xi_0$ is a real constant.
To fit the experimental curve, we have chosen the parameters given for our
simple model and used $T_c=350$~K, which yields the conduction bandwidth of
$2W\simeq 0.49$~eV.
The overall agreement of the calculated curve with the
measured curve should not be taken too literally but the support of the
spontaneous structural phase transition accompanying the exciton condensation in
the mixed-valent system in our theory is noted.
This structural transformation belongs to a well-known class of
{\em ferroelastic} phase transitions \cite{fet}.

By adding a perturbing term to the equations of motion,
we can calculate the first-order changes.  To calculate the
Pauli spin susceptibility, we use the perturbation
\begin{eqnarray}
H_B= -\mu_B \sum_{{\bf k}\sigma}
(g_d\sigma d^{\dagger}_{{\bf k}\sigma}d^{\vphantom\dagger}_{{\bf k}\sigma}
+g_f\sigma f^{\dagger}_{{\bf k}\sigma}f^{\vphantom\dagger}_{{\bf k}\sigma} )B,
\end{eqnarray}
where $B$ is the magnetic field in a fixed direction, $\mu_B$ is the Bohr
magneton, and $g_f$ and $g_d$ are phenomenological gyromagnetic ratios for $f$
and $d$ electrons. The Pauli susceptibility is
\begin{eqnarray}
\chi&=&\mu_B^2 \sum_{\bf k} |2aV({\bf k})|^2(g_f-g_d)^2
{[f(E_{{\bf k}-})-f(E_{{\bf k}+})]\over (2E_{\bf k})^3}\nonumber\\
&&+(2\mu_B^2/k_BT) \sum_{{\bf k},s=\pm} g_s^2 f(E_{{\bf k}s})
[1-f(E_{{\bf k}s})],
\label{pauli}
\end{eqnarray}
where $g_{\pm}=g_d A_{\mp}({\bf k})+g_f A_{\pm}({\bf k})$ is the effective
gyromagnetic ratio for quasiparticles in the mixed ($\pm$) bands with spectral
weights
$A_{\pm}({\bf k})= \case{1}{2}\{ 1 \mp \xi^{(0)}({\bf k})/E_{\bf k} \}$.
The first sum at the right-hand-side of Eq.~(\ref{pauli}) is due to
interband mixing proportional to $a^2$ of the slave-boson condensate, while the
second sum is from the individual band contributions.  Figure~2 and its inset
show the temperature dependence of the spin susceptibility \cite{Riseborough}
and its relation with
the lattice deformation as well as comparisons with the experimental results for
Ce$_3$Bi$_4$Pt$_3$ \cite{Fisk,AeppliFisk}. The observed low-temperature
upturn of
the susceptibility was sample dependent and was ascribed to magnetic impurities.
Excluding that feature, the calculated temperature dependence of the spin
susceptibility resembles the observation. Fisk {\em et al.} \cite{Fisk} found
an empirical correlation between the lattice distortion and the spin
susceptibility, which is intriguing since $\chi T$ measures the effective
moment squared of an isolated magnetic ion.  Our theory of this functional
correlation provides an alternative explanation in which both properties are
driven by the same condensation phenomenon. Moreover, the theory predicts a
low-temperature indirect gap of
$[a(0)V_0]^2/W\simeq 44$~K, in good agreement with the experimental activation
energy of 42~K determined from transport measurement \cite{Hundley}.
\begin{figure} [!h]
\centering
\leavevmode
\epsfxsize=7cm
\epsfysize=7cm
\epsfbox[18 144 592 718] {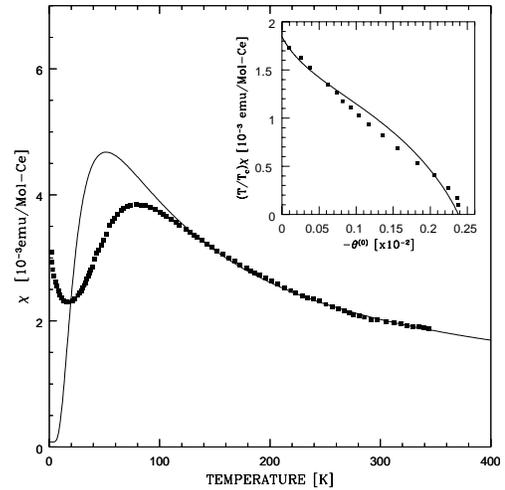}
\caption[]
{\label{fig2} Temperature dependence of static Pauli spin susceptibility
and its correlation with lattice deformation (inset) for
$g_d = 2.38$, $g_f= 1.55$ for the same model as in Fig.~1.
Squares are experimental results from Ref.~10
for Ce$_3$Bi$_4$Pt$_3$.}
\end{figure}

Since the exciton condensation is intimately connected with the
structural transformation, one would expect corresponding changes in elastic
constants and phonon properties.  With the addition of the deformation terms in
Eqs.~(\ref{wvfn}) and (\ref{boso}), to first order in
the external stress in Eq.~(\ref{deform}), the dynamics of the pseudospin and
the boson condensate is governed by a set of five coupled linear equations.
The linear response deduced from them involves the
collective mode in general \cite{Portlong}.
The elastic constant is
$C+C_{\text{cond}}$, with the second being from the response of the
condensate.  At the high-frequency sound wave limit, where $\omega\gg \Gamma$
(the linewidth of  the collective mode, taken to be of the order of the indirect
gap) but $\omega\ll 2aV_0$ (the direct gap) ,
the response involves the collective mode and the dynamic elastic constant is

\begin{eqnarray}
C_{\text{cond}} & & = 4 a^2 N_{\rm s} \Xi_0^2 \times \nonumber \\
& &\frac{4 D_2-4 \lambda D_1 + \lambda^{2} D_0
+24 \lambda V_0^2 (D_0 D_2-D_1^2)}
{(1+4 V_0^2 D_1)^2-V_0^2 D_0 (16 V_0^2 D_2-2 \lambda)} ,
\end{eqnarray}
where
$\lambda=\lambda^{(0)}$
for short, and
\begin{eqnarray}
D_i(\omega)=\frac{1}{N_{\rm s}}\sum_{\bf k} [\xi^{(0)}({\bf k})]^i \;
\frac{ f(E_{{\bf k}-}) - f(E_{{\bf k}+})}
{2E_{\bf k}(\omega^2-4E_{\bf k}^2)}.
\end{eqnarray}
The relative velocity change $\Delta v_s/v_s = \Re(C_{cond})/2C$  is plotted in
the inset of Fig.~3 for our simple model.  For the parameters of Fig.~1,
the unit
of relative change is $N_{\rm s}\Xi_0^2/CW=0.26\%$.  It is evident that its
temperature
dependence reflects the second-order phase transition and differs greatly from
the typical anharmonic effect.
In the static or low-frequency limit ($\omega\ll \Gamma$), the elastic
response is decoupled from the collective mode and is given by the
quasi-particle
contributions.  There is a characteristic step-like drop
\cite{fet} in the inverse static elastic constant at
$T_c$, with the magnitude of the drop proportional to $\Xi_0^2$.  The static
elastic constant is qualitatively different from the high-frequency one.
\begin{figure} [!b]
\centering
\leavevmode
\epsfxsize=7cm
\epsfysize=7cm
\epsfbox[18 144 592 718] {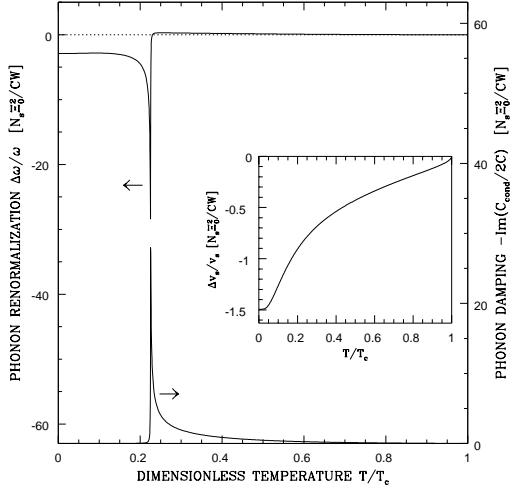}
\caption[]
{\label{fig3} Relative phonon energy change and intrinsic linewidth in units of
the dimensionless factor $N_{\rm s}\Xi_0^2/CW$
for a phonon of unrenormalized frequency $\omega/W =0.2 +i 10^{-4}$ for the
simple model described in the text. Inset:
Relative velocity change for the case of lattice softening for ultrasound of
$\omega/W=10^{-6}$.}
\end{figure}

In the study of high-frequency phonons, we have taken, for simplicity, the
electron-phonon interaction to be given by the same  deformation potential as
above.  A phonon, with energy greater  than the direct gap
($\omega > 2aV_0$)  but limited by the Debye frequency
 ($\sim 0.1$~eV),
can break particle-hole pairs which contribute to the
imaginary part of $C_{cond}$. This is the well-known Landau damping.
The relative phonon linewidth contains a term $-\Im[ C_{\text{cond}}]/2C$.
Correspondingly, the relative phonon energy change is given by
$\Re [C_{\text{cond}}]/2C$.  Figure~3 presents the numerical results for such a
high-energy phonon within our simple model.  The resonance behavior in both the
frequency change and damping occurs when the phonon frequency equals the
temperature-dependent direct gap. Neutron scattering by phonons
\cite{neutron} should be able to detect such resonances.  Similar behavior in
superconductors has been predicted \cite{allen} and observed \cite{neutron}.

We have also considered a two-dimensional square lattice with unit cell
size $d$ and a conduction band $\epsilon ({\bf k})=$
$-\frac{W}{2}(\cos k_xd + \cos k_yd)$. For hybridization $V({\bf k})=$
$iV_0\sin (k_xd/2)$ between a $d$-electron at the corner and an $f$
in between, the ground state is a fully gapped insulator, similar to
our simple model presented above. For hybridization $V({\bf k})=$
$iV_0\sin (k_xd)$ between d-f electrons at neighboring corners the exciton
condensate contains nodes in the excitation spectrum. Nevertheless
the numerical results for $a(T)$ and consequently the lattice deformation
resemble those in Fig.1. Its interesting low-temperature transport
properties are beyond the scope of this Letter.

In summary, the additional consideration of the rotational symmetry of the
hybridization between $d$ and $f$ electrons to the finite-temperature,
mean-field
solution of the periodic Anderson lattice using the slave-boson technique
reveals the correlated insulator state to be a $p$-wave exciton condensate
as well as the existence of a concomitant structural transformation.
Theoretical deduction of the temperature dependence of the spontaneous lattice
distortion and of the Pauli spin susceptibility have some experimental
support.  In particular, the empirical correlation \cite{Fisk}
between the lattice distortion and the spin susceptibility in the correlated
insulator Ce$_3$Bi$_4$Pt$_3$ can be explained in terms of the temperature
dependence of exciton condensation.  Moreover, intimate relation between the
symmetry of the exciton condensate and the lattice distortion leads to
theoretical predictions of (1) characteristic static elastic
constant jump across the phase transition, (2) temperature dependence of the
high frequency ultrasound velocity, and (3) resonance behavior of phonon
frequency and damping as the gap parameter varies with temperature. All these
consequences of the theory could be easily subjected to further experimental
tests in mixed-valent correlated insulators.

We thank M. B. Maple, S. H. Liu, and S. R. Renn for helpful discussions.
JMD also thanks R. C. Dynes and A. J. Leggett for
discussions and encouragement.
This work was supported by NSF DMR 91-13631 (JMD and DPA),
and by DMR 94-21966 (LJS).


\vskip -0.2cm

\begin{references}

\bibitem{AeppliFisk} G. Aeppli and Z. Fisk, {\sl Comm. Condens. Matter Phys.}
{\bf 16}, 155 (1992).

\bibitem{GangOf5} P. A. Lee, {\em et al.}, {\sl Comm. Condens. Matter Phys.}
{\bf 12}, 99 (1986).

\bibitem{Varma} C. M. Varma, {\sl Phys. Rev. B} {\bf 50}, 9952 (1994).

\bibitem{Millis}  A. Auerbach and K. Levin, {\sl Phys. Rev. Lett.} {\bf 57},
877 (1986); A. J. Millis and P.A. Lee, {\sl Phys. Rev. B} {\bf 35}, 3394 (1987).

\bibitem{Riseborough} P. S. Riseborough, {\sl Phys. Rev. B}
{\bf 45}, 13984 (1992); A. J. Millis in {\sl Physical Phenomena at High
Magnetic Fields} (ed.\ E. Manousakis {\it et al.\/}), p. 146 (Addison-Wesley,
Redwood City, 1992).


\bibitem{fet} A. Bulou, M. Rousseau and J. Nouet, {\sl Key Engineering
Materials} {\bf 68}, 133 (1992).

\bibitem{Portengen} T. Portengen, Th. \"{O}streich, and L. J. Sham,
{\sl Phys. Rev. Lett.} {\bf 76}, 3384 (1996).

\bibitem{foot} A nonzero expectation value for $b$
would violate a local gauge symmetry.  Higher orders of perturbation
theory beyond the mean-field level will ensure $\langle b\rangle=0$
and a power law dependence to the correlator $\langle b(t) b^\dagger(0)\rangle$
[see, {\it e.g.\/} N. Read, {\sl J. Phys. C} {\bf 18}, 2651 (1985)] -- there
is no true Goldstone mode for this condensate, but the insulating
gap, which depends on $b^{\dagger}b$, persists at low temperatures.

\bibitem{ShamZiman} See, for example, L. J. Sham and J. M. Ziman,
{\sl Solid State Phys.} {\bf 15}, 221 (1963).

\bibitem{Fisk} Z. Fisk {\it et al.}, {\sl J. Alloys Comp.}
{\bf 181}, 369 (1992).

\bibitem{Hundley} M. F. Hundley {\it et al.}, {\sl Phys. Rev. B}
{\bf 42}, 6842 (1990).

\bibitem{Portlong} T. Portengen, Th. \"{O}streich, and L. J. Sham,
{\sl Phys. Rev. B}, in press.

\bibitem{neutron} J. D. Axe, {\sl Physica} {\bf 137}B, 107(1986).

\bibitem{allen} V. M. Bobetic, {\sl Phys. Rev.} {\bf 136}, 1535 (1964);
  P. B. Allen, {\sl Phys. Rev. B} {\bf 6}, 2577 (1972).

\end{references}
\end{document}